\documentclass{article}
\usepackage{amssymb}
\usepackage{amsmath}

\begin{document}

\title{New n-mode squeezing operator and squeezed states with standard
squeezing \thanks{{\small Work supported by the National Natural Science
Foundation of China under grants 10775097 and 10874174.}}}
\author{Li-yun Hu$^{1,2}$\thanks{{\small Corresponding author. E-mail
addresses: hlyun2008@126.com or hlyun@sjtu.edu.cn.}} and Hong-yi Fan$^{1}$ \\
$^{1}${\small Department of Physics, Shanghai Jiao Tong University, Shanghai 200030, China}\\
{\small \ }$^{2}${\small College of Physics \& Communication Electronics,
Jiangxi Normal University, Nanchang 330022, China}}
\maketitle

\begin{abstract}
{\small We find that the exponential operator }$V\equiv \exp \left[ \mathtt{i%
}\lambda \left( Q_{1}P_{2}+Q_{2}P_{3}+\cdots +Q_{n-1}P_{n}+Q_{n}P_{1}\right) %
\right] ,${\small \ }$Q_{i},${\small \ }$P_{i}${\small \ are respectively
the coordinate and momentum operators, is an n-mode squeezing operator which
engenders standard squeezing. By virtue of the technique of integration
within an ordered product of operators we derive }$V${\small 's normally
ordered expansion and obtain the n-mode squeezed vacuum states, its Wigner
function is calculated by using the Weyl ordering invariance under similar
transformations.}
\end{abstract}



\section{Introduction}

Quantum entanglement is a weird, remarkable feature of quantum mechanics
though it implies intricacy. In recent years, various entangled states have
brought considerable attention and interests of physicists because of their
potential uses in quantum communication \cite{1,2}. Among them the two-mode
squeezed state exhibits quantum entanglement between the idle-mode and the
signal-mode in a frequency domain manifestly, and is a typical entangled
state of continuous variable. Theoretically, the two-mode squeezed state is
constructed by the two-mode squeezing operator $S=\exp [\lambda
(a_{1}a_{2}-a_{1}^{\dagger }a_{2}^{\dagger })]$ \cite{3,4,5} acting on the
two-mode vacuum state $\left\vert 00\right\rangle $,%
\begin{equation}
S\left\vert 00\right\rangle =\text{sech}\lambda \exp \left[
-a_{1}^{^{\dagger }}a_{2}^{^{\dagger }}\tanh \lambda \right] \left\vert
00\right\rangle ,  \label{1}
\end{equation}%
where $\lambda $ is a squeezing parameter, the disentangling of $S$ can be
obtained by using SU(1,1) Lie algebra, $[a_{1}a_{2},a_{1}^{\dagger
}a_{2}^{\dagger }]=a_{1}^{\dagger }a_{1}+a_{2}^{\dagger }a_{2}+1,$ or by
using the entangled state representation $\left\vert \eta =\eta _{1}+i\eta
_{2}\right\rangle $ \cite{6,7}%
\begin{equation}
\left\vert \eta \right\rangle =\exp \left[ -\frac{1}{2}\left\vert \eta
\right\vert ^{2}+\eta a_{1}^{\dagger }-\eta ^{\ast }a_{2}^{\dagger
}+a_{1}^{\dagger }a_{2}^{\dagger }\right] \left\vert 00\right\rangle ,
\label{2}
\end{equation}%
$\left\vert \eta \right\rangle $ is the common eigenvector of two particles'
relative position $\left( Q_{1}-Q_{2}\right) $ and the tota momentum\ $%
\left( P_{1}+P_{2}\right) $, obeys the eigenvector equation, $\left(
Q_{1}-Q_{2}\right) \left\vert \eta \right\rangle =\sqrt{2}\eta
_{1}\left\vert \eta \right\rangle ,$ $\left( P_{1}+P_{2}\right) =\left\vert
\eta \right\rangle =\sqrt{2}\eta _{2}\left\vert \eta \right\rangle ,$ and
the orthonormal-complete relation%
\begin{equation}
\int \frac{d^{2}\eta }{\pi }\left\vert \eta \right\rangle \left\langle \eta
\right\vert =1,\text{ }\left\langle \eta ^{\prime }\right\vert \left. \eta
\right\rangle =\pi \delta \left( \eta -\eta ^{\prime }\right) \left( \eta
^{\ast }-\eta ^{\prime \ast }\right) ,  \label{4}
\end{equation}%
because the two-mode squeezing operator has its natural representation in $%
\left\langle \eta \right\vert $ basis%
\begin{equation}
S=\exp \left[ \lambda \left( a_{1}a_{2}-a_{1}^{\dagger }a_{2}^{\dagger
}\right) \right] =\int \frac{d^{2}\eta }{\pi \mu }\left\vert \frac{\eta }{%
\mu }\right\rangle \left\langle \eta \right\vert ,\text{ }S\left\vert \eta
\right\rangle =\frac{1}{\mu }\left\vert \frac{\eta }{\mu }\right\rangle ,%
\text{ }\mu =e^{\lambda },  \label{5}
\end{equation}%
The proof of Eq.(\ref{5}) is proceeded by virtue of the technique of
integration within an ordered product (IWOP) of operators \cite{8,9,10}%
\begin{eqnarray}
\int \frac{d^{2}\eta }{\pi \mu }\left\vert \eta /\mu \right\rangle
\left\langle \eta \right\vert  &=&\int \frac{d^{2}\eta }{\pi \mu }\colon
\exp \left\{ -\frac{\mu ^{2}+1}{2\mu ^{2}}|\eta |^{2}+\eta \left( \frac{%
a_{1}^{^{\dagger }}}{\mu }-a_{2}\right) \right.   \notag \\
&&\left. +\eta ^{\ast }\left( a_{1}-\frac{a_{2}^{^{\dagger }}}{\mu }\right)
+a_{1}^{\dagger }a_{2}^{^{\dagger }}+a_{1}a_{2}-a_{1}^{\dagger
}a_{1}-a_{2}^{\dagger }a_{2}\right\} \colon   \notag \\
&=&\frac{2\mu }{1+\mu ^{2}}\colon \exp \left\{ \frac{\mu ^{2}}{1+\mu ^{2}}%
\left( \frac{a_{1}^{^{\dagger }}}{\mu }-a_{2}\right) \left( a_{1}-\frac{%
a_{2}^{^{\dagger }}}{\mu }\right) -\left( a_{1}-a_{2}^{^{\dagger }}\right)
\left( a_{1}^{^{\dagger }}-a_{2}\right) \right\} \colon   \notag \\
&=&e^{-a_{1}^{^{\dagger }}a_{2}^{^{\dagger }}\tanh \lambda
}e^{(a_{1}^{^{\dagger }}a_{1}+a_{2}^{^{\dagger }}a_{2}+1)\ln \text{sech}%
\lambda }e^{a_{1}a_{2}\tanh \lambda }\equiv S,  \label{6}
\end{eqnarray}%
Eq. (\ref{5}) confirms that the two-mode squeezed state itself is an
entangled state which entangles the idle mode and signal mode as an outcome
of a parametric-down conversion process \cite{11}. The $\left\vert \eta
\right\rangle $ state was constructed in Ref. \cite{6,7} according to the
idea of Einstein, Podolsky and Rosen in their argument that quantum
mechanics is incomplete \cite{12}.

Using the relation between bosonic operators and the coordinate $Q_{i},$
momentum $P_{i},$ $Q_{i}=( a_{i}+a_{i}^{\dagger }) /\sqrt{2},\ P_{i}=(
a_{i}-a_{i}^{\dagger }) /(\sqrt{2}\mathtt{i}),$ and introducing the two-mode
quadrature operators of light field as in Ref. \cite{4}, $%
x_{1}=(Q_{1}+Q_{2})/2,x_{2}=(P_{1}+P_{2})/2,$ the variances of $x_{1}$ and $%
x_{2}$ in the state $S\left\vert 00\right\rangle $ are in the standard form%
\begin{equation}
\left\langle 00\right\vert S^{\dagger }x_{2}^{2}S\left\vert 00\right\rangle =%
\frac{1}{4}e^{-2\lambda },\text{ \ }\left\langle 00\right\vert S^{\dagger
}x_{1}^{2}S\left\vert 00\right\rangle =\frac{1}{4}e^{2\lambda },  \label{7}
\end{equation}%
thus we get the standard squeezing for the two quadrature: $x_{1}\rightarrow
\frac{1}{2}e^{\lambda }x_{1},$ $x_{2}\rightarrow \frac{1}{2}e^{-\lambda
}x_{2}$. On the other hand, the two-mode squeezing operator can also be
recast into the form $S=\exp \left[ \mathtt{i}\lambda \left(
Q_{1}P_{2}+Q_{2}P_{1}\right) \right] .$ Then an interesting question
naturally rises: what is the property of the $n$-mode operator
\begin{equation}
V\equiv \exp \left[ \mathtt{i}\lambda \left( Q_{1}P_{2}+Q_{2}P_{3}+\cdots
+Q_{n-1}P_{n}+Q_{n}P_{1}\right) \right] ,  \label{8}
\end{equation}%
and is it a squeezing operator which can engenders the standard squeezing
for $n$-mode quadratures? What is the normally ordered expansion of $V$ and
what is the state $V\left\vert \mathbf{0}\right\rangle $ ($\left\vert
\mathbf{0}\right\rangle $ is the n-mode vacuum state)? In this work we shall
study $V$ in detail. But how to disentangling the exponential of $V?$ Since
all terms of the set $Q_{i}P_{i+1}\ $($i=1\cdots n$) do not make up a closed
Lie algebra, the problem of what is $V^{\prime }$s the normally ordered form
seems difficult. Thus we appeal to the IWOP technique to solve this problem.
Our work is arranged in this way: firstly we use the IWOP technique to
derive the normally ordered expansion of $V$ and obtain the explicit form of
$\ V\left\vert \mathbf{0}\right\rangle $; then we examine the variances of
the $n$-mode quadrature operators in the state $V\left\vert \mathbf{0}%
\right\rangle $, we find that $V$ just causes standard squeezing. Thus $V$
is a squeezing operator. The Wigner function of $V\left\vert \mathbf{0}%
\right\rangle $ is calculated by using the Weyl ordering invariance under
similar transformations. Some examples are discussed in the last section.

\section{The normal product form of $V$}

In order to disentangle operator $V$, let $A$ be%
\begin{equation}
A=\left(
\begin{array}{ccccc}
0 & 1 & 0 & \cdots  & 0 \\
0 & 0 & 1 & \cdots  & 0 \\
\vdots  & \vdots  & \ddots  & \ddots  & 0 \\
0 & 0 & \cdots  & \ddots  & 1 \\
1 & 0 & \cdots  & \cdots  & 0%
\end{array}%
\right) ,  \label{9}
\end{equation}%
then $V$ in (\ref{8}) is compactly expressed as%
\begin{equation}
V=\exp \left[ \mathtt{i}\lambda \underset{i,j=1}{\overset{n}{\sum }}%
Q_{i}A_{ij}P_{j}\right] .  \label{10}
\end{equation}%
Using the Baker-Hausdorff formula, $e^{A}Be^{-A}=B+\left[ A,B\right] +\frac{1%
}{2!}\left[ A,\left[ A,B\right] \right] +\frac{1}{3!}\left[ A,\left[ A,\left[
A,B\right] \right] \right] +\cdots ,$we have $($here and henceforth the
repeated indices represent the Einstein summation notation)%
\begin{eqnarray}
V^{-1}Q_{k}V &=&Q_{k}-\lambda Q_{i}A_{ik}+\frac{1}{2!}\mathtt{i}\lambda ^{2}%
\left[ Q_{i}A_{ij}P_{j},Q_{l}A_{lk}\right] +...  \notag \\
&=&Q_{i}(e^{-\lambda A})_{ik}=(e^{-\lambda \tilde{A}})_{ki}Q_{i},  \label{12}
\\
V^{-1}P_{k}V &=&P_{k}+\lambda A_{ki}P_{i}+\frac{1}{2!}\mathtt{i}\lambda ^{2}%
\left[ A_{ki}P_{j},Q_{l}A_{lm}P_{m}\right] +...  \notag \\
&=&(e^{\lambda A})_{ki}P_{i},  \label{12a}
\end{eqnarray}%
From Eq.(\ref{12}) we see that when $V$ acts on the n-mode coordinate
eigenstate $\left\vert \vec{q}\right\rangle ,$ where $\widetilde{\vec{q}}%
=(q_{1},q_{2},\cdots ,q_{n})$, it squeezes $\left\vert \vec{q}\right\rangle $
in the way of
\begin{equation}
V\left\vert \vec{q}\right\rangle =\left\vert \Lambda \right\vert
^{1/2}\left\vert \Lambda \vec{q}\right\rangle ,\text{ }\Lambda =e^{-\lambda
\tilde{A}},\text{ }\left\vert \Lambda \right\vert \equiv \det \Lambda .
\label{13}
\end{equation}%
Thus $V$ has the representation on the coordinate $\left\langle \vec{q}%
\right\vert $ basis
\begin{equation}
V=\int d^{n}qV\left\vert \vec{q}\right\rangle \left\langle \vec{q}%
\right\vert =\left\vert \Lambda \right\vert ^{1/2}\int d^{n}q\left\vert
\Lambda \vec{q}\right\rangle \left\langle \vec{q}\right\vert ,\text{ \ \ }%
V^{\dagger }=V^{-1},  \label{14}
\end{equation}%
since $\int d^{n}q\left\vert \vec{q}\right\rangle \left\langle \vec{q}%
\right\vert =1.$ Using the expression of eigenstate $\left\vert \vec{q}%
\right\rangle $ in Fock space
\begin{eqnarray}
&\left\vert \vec{q}\right\rangle =\pi ^{-n/4}\colon \exp [-\frac{1}{2}%
\widetilde{\vec{q}}\vec{q}+\sqrt{2}\widetilde{\vec{q}}a^{\dag }-\frac{1}{2}%
\tilde{a}^{\dag }a^{\dag }]\left\vert \mathbf{0}\right\rangle ,\text{ }&
\label{15} \\
&\tilde{a}^{\dag }=(a_{1}^{\dag },a_{2}^{\dag },\cdots ,a_{n}^{\dag })\text{,%
}&  \notag
\end{eqnarray}%
and $\left\vert \mathbf{0}\right\rangle \left\langle \mathbf{0}\right\vert
=\colon \exp [-\tilde{a}^{\dag }a^{\dag }]\colon ,$ we can put $V$ into the
normal ordering form ,
\begin{eqnarray}
V &=&\pi ^{-n/2}\left\vert \Lambda \right\vert ^{1/2}\int d^{n}q\colon \exp
[-\frac{1}{2}\widetilde{\vec{q}}(1+\widetilde{\Lambda }\Lambda )\vec{q}+%
\sqrt{2}\widetilde{\vec{q}}(\widetilde{\Lambda }a^{\dag }+a)  \notag \\
&&-\frac{1}{2}(\widetilde{a}a+\tilde{a}^{\dag }a^{\dag })-\tilde{a}^{\dag
}a]\colon .  \label{16}
\end{eqnarray}%
To compute the integration in Eq.(\ref{16}) by virtue of the IWOP technique,
we use the mathematical formula
\begin{equation}
\int d^{n}x\exp [-\widetilde{x}Fx+\widetilde{x}v]=\pi ^{n/2}(\det
F)^{-1/2}\exp \left[ \frac{1}{4}\widetilde{v}F^{-1}v\right] ,  \label{17}
\end{equation}%
then we derive
\begin{eqnarray}
V &=&\left( \frac{\det \Lambda }{\det N}\right) ^{1/2}\exp \left[ \frac{1}{2}%
\tilde{a}^{\dag }\left( \Lambda N^{-1}\widetilde{\Lambda }-I\right) a^{\dag }%
\right]   \notag \\
&&\times \colon \exp \left[ \tilde{a}^{\dag }\left( \Lambda N^{-1}-I\right) a%
\right] \colon \exp \left[ \frac{1}{2}\widetilde{a}\left( N^{-1}-I\right) a%
\right] ,  \label{18}
\end{eqnarray}%
where $N=(1+\widetilde{\Lambda }\Lambda )/2$. Eq.(\ref{18}) is just the
normal product form of $V.$

\section{The squeezing property of $V\left\vert \mathbf{0}\right\rangle $}

Operating $V$ on the n-mode vacuum state $\left\vert \mathbf{0}\right\rangle
,$ we obtain the squeezed vacuum state%
\begin{equation}
V\left\vert \mathbf{0}\right\rangle =\left( \frac{\det \Lambda }{\det N}%
\right) ^{1/2}\exp \left[ \frac{1}{2}\tilde{a}^{\dag }\left( \Lambda N^{-1}%
\widetilde{\Lambda }-I\right) a^{\dag }\right] \left\vert \mathbf{0}%
\right\rangle .  \label{19}
\end{equation}%
Now we evaluate the variances of the n-mode quadratures. The quadratures in
the n-mode case are defined as%
\begin{equation}
X_{1}=\frac{1}{\sqrt{2n}}\sum_{i=1}^{n}Q_{i},\text{ }X_{2}=\frac{1}{\sqrt{2n}%
}\sum_{i=1}^{n}P_{i},  \label{20}
\end{equation}%
obeying $[X_{1},X_{2}]=\frac{\mathtt{i}}{2}.$ Their variances are $\left(
\Delta X_{i}\right) ^{2}=\left\langle X_{i}^{2}\right\rangle -\left\langle
X_{i}\right\rangle ^{2}$, $i=1,2.$ Noting the expectation values of $X_{1}$
and $X_{2}$ in the state $V\left\vert \mathbf{0}\right\rangle $, $%
\left\langle X_{1}\right\rangle =\left\langle X_{2}\right\rangle =0,$ and
using Eqs. (\ref{12}) and (\ref{12a}) we see that the variances are%
\begin{eqnarray}
\left( \triangle X_{1}\right) ^{2} &=&\left\langle \mathbf{0}\right\vert
V^{-1}X_{1}^{2}V\left\vert \mathbf{0}\right\rangle =\frac{1}{2n}\left\langle
\mathbf{0}\right\vert
V^{-1}\sum_{i=1}^{n}Q_{i}\sum_{j=1}^{n}Q_{j}V\left\vert \mathbf{0}%
\right\rangle   \notag \\
&=&\frac{1}{2n}\left\langle \mathbf{0}\right\vert
\sum_{i=1}^{n}Q_{k}(e^{-\lambda A})_{ki}\sum_{j=1}^{n}(e^{-\lambda \tilde{A}%
})_{jl}Q_{l}\left\vert \mathbf{0}\right\rangle   \notag \\
&=&\frac{1}{2n}\underset{i,j}{\sum^{n}}(e^{-\lambda A})_{ki}(e^{-\lambda
\tilde{A}})_{jl}\left\langle \mathbf{0}\right\vert Q_{k}Q_{l}\left\vert
\mathbf{0}\right\rangle   \notag \\
&=&\frac{1}{4n}\underset{i,j}{\sum^{n}}(e^{-\lambda A})_{ki}(e^{-\lambda
\tilde{A}})_{jl}\left\langle \mathbf{0}\right\vert a_{k}a_{l}^{\dagger
}\left\vert \mathbf{0}\right\rangle   \notag \\
&=&\frac{1}{4n}\underset{i,j}{\sum^{n}}(e^{-\lambda A})_{ki}(e^{-\lambda
\tilde{A}})_{jl}\delta _{kl}=\frac{1}{4n}\underset{i,j}{\sum^{n}}(\widetilde{%
\Lambda }\Lambda )_{ij},  \label{22}
\end{eqnarray}%
similarly we have%
\begin{equation}
\left( \triangle X_{2}\right) ^{2}=\left\langle \mathbf{0}\right\vert
V^{-1}X_{2}^{2}V\left\vert \mathbf{0}\right\rangle =\frac{1}{4n}\underset{i,j%
}{\sum^{n}}(\widetilde{\Lambda }\Lambda )_{ij}^{-1},  \label{22a}
\end{equation}%
Eqs. (\ref{22}) -(\ref{22a}) are the quadrature variance formula in the
transformed vacuum state acted by the operator $\exp [\mathtt{i}\lambda
\underset{i,j=1}{\overset{n}{\sum }}Q_{i}A_{ij}P_{j}].$ By observing that $A$
in (\ref{9}) is a cyclic matrix, we see%
\begin{equation}
\underset{i,j}{\sum^{n}}\left[ (A+\tilde{A})^{l}\right] _{i\text{ }j}=2^{l}n,
\label{22b}
\end{equation}%
then using $A\tilde{A}=\tilde{A}A,$ so $\widetilde{\Lambda }\Lambda
=e^{-\lambda (A+\tilde{A})}$, a symmetric matrix, we have
\begin{equation}
\underset{i,j=1}{\sum^{n}}(\widetilde{\Lambda }\Lambda )_{i\text{ }%
j}=\sum_{l=0}^{\infty }\frac{(-\lambda )^{l}}{l!}\underset{i,j}{\sum^{n}}%
\left[ (A+\tilde{A})^{l}\right] _{i\text{ }j}=n\sum_{l=0}^{\infty }\frac{%
(-\lambda )^{l}}{l!}2^{l}=ne^{-2\lambda },  \label{21}
\end{equation}%
and%
\begin{equation}
\underset{i,j=1}{\sum^{n}}(\widetilde{\Lambda }\Lambda )_{i\text{ }%
j}^{-1}=ne^{2\lambda }.  \label{21a}
\end{equation}%
it then follows%
\begin{eqnarray}
\left( \triangle X_{1}\right) ^{2} &=&\frac{1}{4n}\underset{i,j}{\sum^{n}}(%
\widetilde{\Lambda }\Lambda )_{ij}=\frac{e^{-2\lambda }}{4},  \label{21b} \\
\left( \triangle X_{2}\right) ^{2} &=&\frac{1}{4n}\underset{i,j}{\sum^{n}}(%
\widetilde{\Lambda }\Lambda )_{ij}^{-1}=\frac{e^{2\lambda }}{4}.  \label{21c}
\end{eqnarray}%
This leads to $\triangle X_{1}\cdot \triangle X_{2}=\frac{1}{4},$ which
shows that $V$ is a correct n-mode squeezing operator for the n-mode
quadratures in Eq.(\ref{20}) and produces the standard squeezing similar to
Eq. (\ref{7}).

\section{The Wigner function of $V\left\vert \mathbf{0}\right\rangle $}

Wigner distribution functions \cite{13,14,15} of quantum states are widely
studied in quantum statistics and quantum optics. Now we derive the
expression of the Wigner function of $V\left\vert \mathbf{0}\right\rangle .$
Here we take a new method to do it. Recalling that in Ref.\cite{16,17,18} we
have introduced the Weyl ordering form of single-mode Wigner operator $%
\Delta \left( q,p\right) $,%
\begin{equation}
\Delta _{1}\left( q_{1},p_{1}\right)
=\genfrac{}{}{0pt}{}{:}{:}\delta \left( q_{1}-Q_{1}\right) \delta
\left( p_{1}-P_{1}\right) \genfrac{}{}{0pt}{}{:}{:}, \label{23a}
\end{equation}%
its normal ordering form is%
\begin{equation}
\Delta _{1}\left( q_{1},p_{1}\right) =\frac{1}{\pi }\colon \exp \left[
-\left( q_{1}-Q_{1}\right) ^{2}-\left( p_{1}-P_{1}\right) ^{2}\right] \colon
\label{23}
\end{equation}%
where the symbols $\colon \colon $ and
$\genfrac{}{}{0pt}{}{:}{:}\genfrac{}{}{0pt}{}{:}{:}$ denote the
normal ordering and the Weyl ordering, respectively. Note that the
order of Bose operators $a_{1}$ and $a_{1}^{\dagger }$ within a
normally ordered product and a Weyl ordered product can be permuted.
That is to say, even though $[a_{1},a_{1}^{\dagger }]=1$, we can
have $\colon a_{1}a_{1}^{\dagger
}\colon =\colon a_{1}^{\dagger }a_{1}\colon $ and$\genfrac{}{}{0pt}{}{:}{:}%
a_{1}a_{1}^{\dagger }\genfrac{}{}{0pt}{}{:}{:}=\genfrac{}{}{0pt}{}{:}{:}a_{1}^{\dagger }a_{1}\genfrac{}{}{0pt}{}{:}{%
:}.$ The Weyl ordering has a remarkable property, i.e., the
order-invariance of Weyl ordered operators under similar
transformations \cite{16,17,18},
which means%
\begin{equation}
U\genfrac{}{}{0pt}{}{:}{:}\left( \circ \circ \circ \right) \genfrac{}{}{0pt}{}{:}{:}U^{-1}=\genfrac{}{}{0pt}{}{:}{:}%
U\left( \circ \circ \circ \right) U^{-1}\genfrac{}{}{0pt}{}{:}{:},
\label{24}
\end{equation}%
as if the \textquotedblleft fence"
$\genfrac{}{}{0pt}{}{:}{:}\genfrac{}{}{0pt}{}{:}{:}$did not exist.

For n-mode case, the Weyl ordering form of the Wigner operator is
\begin{equation}
\Delta _{n}\left( \vec{q},\vec{p}\right) =\genfrac{}{}{0pt}{}{:}{:}\delta \left( \vec{q}-%
\vec{Q}\right) \delta \left( \vec{p}-\vec{P}\right)
\genfrac{}{}{0pt}{}{:}{:}, \label{25}
\end{equation}%
where $\widetilde{\vec{Q}}=(Q_{1},Q_{2},\cdots ,Q_{n})$ and $\widetilde{\vec{%
P}}=(P_{1},P_{2},\cdots ,P_{n})$. Then according to the Weyl ordering
invariance under similar transformations and Eqs.(\ref{12}) and (\ref{12a})
we have%
\begin{eqnarray}
V^{-1}\Delta _{n}\left( \vec{q},\vec{p}\right) V
&=&V^{-1}\genfrac{}{}{0pt}{}{:}{:}\delta
\left( \vec{q}-\vec{Q}\right) \delta \left( \vec{p}-\vec{P}\right) \genfrac{}{}{0pt}{}{:}{%
:}V  \notag \\
&=&\genfrac{}{}{0pt}{}{:}{:}\delta \left( q_{k}-(e^{-\lambda
\tilde{A}})_{ki}Q_{i}\right)
\delta \left( p_{k}-(e^{rA})_{ki}P_{i}\right) \genfrac{}{}{0pt}{}{:}{:}  \notag \\
&=&\genfrac{}{}{0pt}{}{:}{:}\delta \left(
e^{r\tilde{A}}\vec{q}-\vec{Q}\right) \delta
\left( e^{-rA}\vec{p}-\vec{P}\right) \genfrac{}{}{0pt}{}{:}{:}  \notag \\
&=&\Delta \left( e^{r\tilde{A}}\vec{q},e^{-rA}\vec{p}\right) ,
\label{26}
\end{eqnarray}%
Thus using Eqs.(\ref{23a}) and (\ref{26}) the Wigner function of $%
V\left\vert \mathbf{0}\right\rangle $ is
\begin{eqnarray}
&&\left\langle \mathbf{0}\right\vert V^{-1}\Delta _{n}\left( \vec{q},\vec{p}%
\right) V\left\vert \mathbf{0}\right\rangle  \notag \\
&=&\frac{1}{\pi ^{n}}\left\langle \mathbf{0}\right\vert \colon \exp [-(e^{r%
\tilde{A}}\vec{q}-\vec{Q})^{2}-(e^{-rA}\vec{p}-\vec{P})^{2}]\colon
\left\vert \mathbf{0}\right\rangle  \notag \\
&=&\frac{1}{\pi ^{n}}\exp [-(e^{r\tilde{A}}\vec{q})^{2}-\left( e^{-rA}\vec{p}%
\right) ^{2}]  \notag \\
&=&\frac{1}{\pi ^{n}}\exp \left[ -\widetilde{\vec{q}}e^{rA}e^{r\tilde{A}}%
\vec{q}-\widetilde{\vec{p}}e^{-r\tilde{A}}e^{-rA}\vec{p}\right]  \notag \\
&=&\frac{1}{\pi ^{n}}\exp \left[ -\widetilde{\vec{q}}\left( \Lambda
\widetilde{\Lambda }\right) ^{-1}\vec{q}-\widetilde{\vec{p}}\Lambda
\widetilde{\Lambda }\vec{p}\right] ,  \label{27}
\end{eqnarray}
From Eq.(\ref{27}) we see that once the explicit expression of $\Lambda
\tilde{\Lambda}=\exp [-\lambda (A+\tilde{A})]$ is deduced, the Wigner
function of $V\left\vert \mathbf{0}\right\rangle $ can be calculated.

\section{Some examples of calculating the Wigner function}

Taking $n=2$ as an example, $V_{n=2}$ is the usual two-mode squeezing
operator. The matrix $A=\left(
\begin{array}{cc}
0 & 1 \\
1 & 0%
\end{array}%
\right) ,$ it then follows that
\begin{equation}
\Lambda \tilde{\Lambda}=e^{-\lambda (\tilde{A}+A)}=\left(
\begin{array}{cc}
\cosh 2\lambda & -\sinh 2\lambda \\
-\sinh 2\lambda & \cosh 2\lambda%
\end{array}%
\right) ,  \label{29}
\end{equation}%
and%
\begin{equation}
\left( \Lambda \tilde{\Lambda}\right) ^{-1}=\left(
\begin{array}{cc}
\cosh 2\lambda & \sinh 2\lambda \\
\sinh 2\lambda & \cosh 2\lambda%
\end{array}%
\right) .  \label{30}
\end{equation}%
Substituting Eqs.(\ref{29}) and (\ref{30}) into Eq.(\ref{27}), we have
\begin{equation}
\left\langle 00\right\vert V^{-1}\Delta _{2}\left( \vec{q},\vec{p}\right)
V\left\vert 00\right\rangle =\frac{1}{\pi ^{2}}\exp \left[ -2\left( \alpha
_{1}^{\ast }\alpha _{2}^{\ast }+\alpha _{1}\alpha _{2}\right) \sinh 2\lambda
-2\left( \left\vert \alpha _{1}\right\vert ^{2}+\left\vert \alpha
_{2}\right\vert ^{2}\right) \cosh 2\lambda \allowbreak \right] ,  \label{31}
\end{equation}%
where $\alpha _{i}=\frac{1}{\sqrt{2}}\left( q_{i}+\mathtt{i}p_{i}\right)
,(i=1,2.)$. Eq.(\ref{31}) is just the Wigner function of the usual two-mode
squeezing vacuum state. For $n=3,$ we have
\begin{equation}
\Lambda \tilde{\Lambda}=\allowbreak \left(
\begin{array}{ccc}
u & v & \allowbreak v \\
\allowbreak v & u & \allowbreak v \\
v & v & u%
\end{array}%
\right) ,\text{ }u=\frac{2}{3}e^{\lambda }+\frac{1}{3}e^{-2\lambda },\text{ }%
v=\frac{1}{3}\left( \allowbreak e^{-2\lambda }-e^{\lambda }\right) ,
\label{32}
\end{equation}%
and $\left( \Lambda \tilde{\Lambda}\right) ^{-1}$ is obtained by replacing $%
\lambda $ with $-\lambda $ in $\Lambda \tilde{\Lambda}$. By using Eq.(\ref%
{27}) the Wigner function is%
\begin{eqnarray}
\left\langle \mathbf{0}\right\vert V^{-1}\Delta _{3}\left( \vec{q},\vec{p}%
\right) V\left\vert \mathbf{0}\right\rangle &=&\frac{1}{\pi ^{3}}\exp \left[
-\frac{2}{3}\left( \cosh 2\lambda +2\cosh \lambda \right)
\sum_{i=1}^{3}\left\vert \alpha _{i}\right\vert ^{2}\right]  \notag \\
&&\times \exp \left\{ -\frac{1}{3}\allowbreak \left( \sinh 2\lambda -2\sinh
\lambda \right) \sum_{i=1}^{3}\alpha _{i}^{2}\right.  \notag \\
&&-\left. \frac{2}{3}\sum_{j>i=1}^{3}\left[ \left( \cosh 2\lambda -\cosh
\lambda \right) \alpha _{i}\alpha _{j}^{\ast }+\left( \allowbreak \sinh
\lambda +\sinh 2\lambda \right) \alpha _{i}\alpha _{j}\right] +c.c\right\} .
\label{32a}
\end{eqnarray}

For $n=4$ case we have (see the Appendix)
\begin{equation}
\Lambda \tilde{\Lambda}=\allowbreak \left(
\begin{array}{cccc}
u^{\prime } & w^{\prime } & v^{\prime } & w^{\prime } \\
w^{\prime } & u^{\prime } & w^{\prime } & v^{\prime } \\
v^{\prime } & w^{\prime } & u^{\prime } & w^{\prime } \\
w^{\prime } & v^{\prime } & w^{\prime } & u^{\prime }%
\end{array}%
\right) ,  \label{33}
\end{equation}%
where $u^{\prime }=\cosh ^{2}\lambda ,v^{\prime }=\sinh ^{2}\lambda
,w^{\prime }=-\sinh \lambda \cosh \lambda $. Then substituting Eq.(\ref{33})
into Eq.(\ref{27}) we obtain%
\begin{equation}
\left\langle \mathbf{0}\right\vert V^{-1}\Delta _{4}\left( \vec{q},\vec{p}%
\right) V\left\vert \mathbf{0}\right\rangle =\frac{1}{\pi ^{4}}\exp \left\{
-2\cosh ^{2}\lambda \left[ \sum_{i=1}^{4}\left\vert \alpha _{i}\right\vert
^{2}+\left( M+M^{\ast }\right) \tanh ^{2}\lambda +\left( R^{\ast }+R\right)
\allowbreak \tanh \lambda \right] \right\} ,  \label{34}
\end{equation}%
where $M=\alpha _{1}\alpha _{3}^{\ast }+\alpha _{2}\alpha _{4}^{\ast },$ $%
R=\alpha _{1}\alpha _{2}+\alpha _{1}\alpha _{4}+\alpha _{2}\alpha
_{3}+\alpha _{3}\alpha _{4}.$ This form differs evidently from the Wigner
function of the direct-product of usual two two-mode squeezed states' Wigner
functions (\ref{31}). In addition, using Eq. (\ref{33}) we can check Eqs.(%
\ref{21b}) and (\ref{21c}). Further, using Eq.(\ref{33}) we have%
\begin{equation}
N^{-1}=\frac{1}{2}\allowbreak \left(
\begin{array}{cccc}
2 & \tanh \lambda  & 0 & \tanh \lambda  \\
\tanh \lambda  & 2 & \tanh \lambda  & 0 \\
0 & \tanh \lambda  & 2 & \tanh \lambda  \\
\tanh \lambda  & 0 & \tanh \lambda  & 2%
\end{array}%
\right) ,\text{ }\det N=\cosh ^{2}\lambda .  \label{35}
\end{equation}%
Then substituting Eqs.(\ref{35}) and (A.4) into Eq.(\ref{19}) yields the
four-mode squeezed state,%
\begin{equation}
V\left\vert 0000\right\rangle =\text{sech}\lambda \exp \left[ -\frac{1}{2}%
\left( a_{1}^{\dag }+a_{3}^{\dag }\right) \left( a_{2}^{\dag }+a_{4}^{\dag
}\right) \tanh \lambda \right] \left\vert 0000\right\rangle ,  \label{36}
\end{equation}%
from which one can see that the four-mode squeezed state is not the same as
the direct product of two two-mode squeezed states in Eq.(\ref{1}).

In sum, by virtue of the IWOP technique, we have introduced a kind of an
n-mode squeezing operator $V\equiv \exp \left[ \mathtt{i}\lambda \left(
Q_{1}P_{2}+Q_{2}P_{3}+\cdots +Q_{n-1}P_{n}+Q_{n}P_{1}\right) \right] $,
which engenders standard squeezing for the n-mode quadratures. We have
derived $V$'s normally ordered expansion and obtained the expression of
n-mode squeezed vacuum states and evaluated its Wigner function with the aid
of the Weyl ordering invariance under similar transformations.

\textbf{Appendix: Derivation of Eq.(\ref{33})}

For the completeness of this paper, here we derive analytically Eq.(\ref{33}%
). Noticing, for the case of $n=4,$ $A^{4}=I,$ $I$ is the 4$\times $4 unit
matrix, from the Cayley-Hamilton theorem we know that the expanding form of $%
\exp (-r\tilde{A})$ must be
\begin{equation}
\Lambda =\exp (-\lambda \tilde{A})=c_{0}(\lambda )I+c_{1}(\lambda )\tilde{A}%
+c_{2}(\lambda )\tilde{A}^{2}+c_{3}(\lambda )\tilde{A}^{3}.  \tag{A.1}
\end{equation}%
To determine $c_{j}(\lambda )$ , we take $\tilde{A}$ being $e^{\mathtt{i}%
(j/2)\pi }$ $(j=0,1,2,3)$ respectively, then we have%
\begin{equation}
\left\{
\begin{array}{l}
\exp (-\lambda )=c_{0}(\lambda )+c_{1}(\lambda )+c_{2}(\lambda
)+c_{3}(\lambda ), \\
\exp (-\lambda e^{\mathtt{i}(1/2)\pi })=c_{0}(\lambda )+c_{1}(\lambda )e^{%
\mathtt{i}(1/2)\pi }+c_{2}(\lambda )e^{\mathtt{i}\pi }+c_{3}(\lambda )e^{%
\mathtt{i}(3/2)\pi }, \\
\exp (-\lambda e^{\mathtt{i}\pi })=c_{0}(\lambda )+c_{1}(\lambda )e^{\mathtt{%
i}\pi }+c_{2}(\lambda )e^{\mathtt{i}2\pi }+c_{3}(\lambda )e^{\mathtt{i}3\pi
}, \\
\exp (-\lambda e^{\mathtt{i}(3/2)\pi })=c_{0}(\lambda )+c_{1}(\lambda )e^{%
\mathtt{i}(3/2)\pi }+c_{2}(\lambda )e^{\mathtt{i}(6/2)\pi }+c_{3}(\lambda
)e^{\mathtt{i}(9/2)\pi }.%
\end{array}%
\right.   \tag{A.2}
\end{equation}%
Its solution is

\begin{equation}
\left\{
\begin{array}{l}
c_{0}(\lambda )=\frac{1}{2}\left( \cosh \lambda +\cos \lambda \right)  \\
c_{1}(\lambda )=\frac{1}{2}\left( -\sinh \lambda -\sin \lambda \right)  \\
c_{2}(\lambda )=\frac{1}{2}\left( \cosh \lambda -\cos \lambda \right)  \\
c_{3}(\lambda )=\frac{1}{2}\left( -\sinh \lambda +\sin \lambda \right)
\end{array}%
\right. .  \tag{A.3}
\end{equation}%
It follows that%
\begin{equation}
\Lambda =\left(
\begin{array}{cccc}
c_{0} & c_{3} & c_{2} & c_{1} \\
c_{1} & c_{0} & c_{3} & c_{2} \\
c_{2} & c_{1} & c_{0} & c_{3} \\
c_{3} & c_{2} & c_{1} & c_{0}%
\end{array}%
\right) ,\det \Lambda =1,  \tag{A.4}
\end{equation}%
and
\begin{align}
\widetilde{\Lambda }\Lambda & =\left[ c_{0}(\lambda )I+c_{1}(\lambda
)A+c_{2}(\lambda )A^{2}+c_{3}(\lambda )A^{3}\right] \cdot \left[
c_{0}(\lambda )I+c_{1}(\lambda )\tilde{A}+c_{2}(\lambda )\tilde{A}%
^{2}+c_{3}(\lambda )\tilde{A}^{3}\right]   \notag \\
& =\frac{1}{2}\left(
\begin{array}{cccc}
\allowbreak 2\cosh ^{2}\lambda  & -\sinh 2\lambda  & \allowbreak 2\sinh
^{2}\lambda  & -\sinh 2\lambda  \\
-\sinh 2\lambda  & \allowbreak 2\cosh ^{2}\lambda  & -\sinh 2\lambda  &
\allowbreak 2\sinh ^{2}\lambda  \\
\allowbreak 2\sinh ^{2}\lambda  & -\sinh 2\lambda  & \allowbreak 2\cosh
^{2}\lambda  & -\sinh 2\lambda  \\
-\sinh 2\lambda  & \allowbreak 2\sinh ^{2}\lambda  & -\sinh 2\lambda  &
\allowbreak 2\cosh ^{2}\lambda
\end{array}%
\right) ,  \tag{A.5}
\end{align}%
this is just Eq.(\ref{33}).

\textbf{ACKNOWLEDGEMENT} Work supported by the National Natural Science
Foundation of China under grants 10775097 and 10874174.

\end{document}